\documentclass[twocolumn,showpacs,preprintnumbers,amsmath,amssymb]{revtex4}
\usepackage{epsfig}
\usepackage{graphicx}% Include figure files
\usepackage{dcolumn}% Align table columns on decimal point
\usepackage{bm}% bold math

\def\be{\begin{eqnarray}}
\def\en{\end{eqnarray}}
\def\non{\nonumber}
\def\la{\langle}
\def\ra{\rangle}
\def\pr{{\sl Phys. Rev.}~}
\def\prl{{\sl Phys. Rev. Lett.}~}
\def\pl{{\sl Phys. Lett.}~}
\def\np{{\sl Nucl. Phys.}~}

\def\DESepsf(#1 width #2){\epsfxsize=#2 \epsfbox{#1}}
\begin{document}

%\preprint{APS/123-QED}

%%
\title{\boldmath{$B$ Decays as Spectroscope for Charmed Four-quark States}}
%

%\altaffiliation[Also at ]{Physics Department, XYZ University.}
%Lines break automatically or can be forced with \\
\author{$^{a)}$Hai-Yang Cheng}
% \homepage{http://www.Second.institution.edu/~Charlie.Author}
\author{$^{b)}$Wei-Shu Hou}%
%\email{Second.Author@institution.edu}
\affiliation{%
$^{a)}$Institute of Physics, Academia Sinica, Taipei, Taiwan 115,
Republic of China\\
$^{b)}$Department of Physics, National Taiwan University, Taipei,
Taiwan 10764, Republic of China
}%
%\affiliation{$^{b)}$Department of Physics, Chung Yuan Christian
%University, Chung-Li, Taiwan 32023, Republic of China
%}%

\date{\today}% It is always \today, today,
             %  but any date may be explicitly specified

\begin{abstract}
The $D_s(2320)$ state recently observed by BaBar in the
$D_s^+\pi^0$ channel may be the first of a host of $cq\bar q\bar
q$ four-quark states. We give a phenomenological account of the
masses and decay modes. The isosinglet $D_s(2320)$ state is the
only narrow one, dominated by the observed isospin violating decay
and less than $\sim$ 100 keV in width. All other states are
expected to decay hadronically. Notable resonances are in doubly
charged $D_s^+\pi^+$, $D^+K^+$, wrong pairing $D^+K^-$, and also
$D_s^+K^-$, $D\eta$ channels. We propose $B$ decays as searching
ground for such 4-quark states, which recoil against $\bar
D^{(*)}$ meson from $B$ decay, or $\pi^+$ $\bar D^{(*)}$, $\bar K$
mesons from $\bar B$ decay.
Exotic $qc\bar c\bar q$ charmonia could also be produced, and may
be behind the slow $J/\psi$ bump in inclusive $B\to J/\psi+X$
decay.
\end{abstract}

  %Bottom mesons
%\pacs{ %Valid PACS appear here
%}
% PACS, the Physics and Astronomy
                             % Classification Scheme.
%\keywords{Suggested keywords}%Use showkeys class option if keyword
                              %display desired
\maketitle

\section{Introduction}

The BaBar experiment \cite{BaBarDs0} has recently discovered a new
narrow state, the $D_s^+(2320)$, decaying into $D_s^+\pi^0$.
The width is consistent with experimental resolution. It is
lighter than expected from potential models \cite{QM},
considerably below the observed $D_{s1}^+(2536)$ and
$D_{sJ}^+(2573)$ believed to be the $1^+$ and $2^+$ $P$-wave
states with $j = 3/2$ for the $s$ quark spin-orbit angular
momentum. The $D_s^+\pi^0$ decay angular distribution is flat.
These facts suggest a $J^P = 0^+$ assignment, which forbids
$D_s^+(2320) \to D_s^+\pi\pi$ transitions. Since $D_s^+(2320) \to
D_s^+\pi^0$ would violate isospin if it is a normal $c\bar s$
meson, this is consistent with its narrowness. The BaBar
experiment has also searched for $D_s^+(2320) \to D_s^+\gamma$ and
$D_s^+\gamma\gamma$ electromagnetic decays, which are so far
absent. A possible $D_s^{*+}\pi^0$ state could exist at 2460 MeV,
but is not yet established. It was therefore suggested that either
the 2-quark potential models should be revised, or the
$D_s^+(2320)$ could be a four-quark state.

A light $0^+$ hadron nonet exists and could be interpreted as
4-quark states. The $a_0(980)$ and $f_0(980)$ mesons are often
viewed as $K\bar K$ molecules, which accounts for their near
degeneracy with $2m_K$. However, the recent observation of $B^-\to
f_0(980)K^-$ decay \cite{Bellef0,BaBarf0} (and possibly $B^-\to
a_0(980)K^-$) casts doubt to this picture, since it is hard to
conceive a loosely bound state to be ejected in $B$ decay with 2.5
GeV energy. Furthermore, two experiments, E791 at Fermilab
Tevatron \cite{E791}, and BES at BEPC \cite{BES}, have claimed the
possible existence of a scalar resonance, $\kappa$, in the $K\pi$
channel. Together with the resurrected $\sigma$ ($f_0(600)$), one
has an isospin triplet, two doublets and two singlets. They could
be composed of $qq\bar q\bar q$ with $q = u$, $d$, $s$, as we
shall see.

It has been argued \cite{Close} that a strong attraction between
$(qq)_{\bf 3^*}$ and $(\bar q\bar q)_{\bf 3}$ \cite{Jaffe,Alford},
where ${\bf 3^*}$ and {\bf 3} here refer to color, and the absence
of the orbital angular momentum barrier in the $S$-wave 4-quark
state, may explain why the scalar nonet formed by $\sigma$,
$\kappa$, $f_0(980)$ and $a_0(980)$ is lighter than the
conventional $q\bar q$ nonet composed of $f_0(1370)$, $a_0(1450)$,
$K_0^*(1430)$ and $f_0(1500)/f_0(1710)$. By the same token, it is
likely that a scalar $cn\bar n\bar s$ 4-quark state, where $n =
u,\ d$, will be lighter than the $0^+$ $P$-wave $c\bar s$ state,
where a typical potential model prediction gives
2487~MeV~\cite{QM}. Hence, the $cn\bar n\bar s$ state could lie
below the $DK$ threshold and decay only to $D_s^+\pi$ final state.
One is thus motivated to consider the $cq\bar q\bar q$ 4-quark
meson scenario, originally suggested by Lipkin \cite{Lipkin}, and
explore its possible multiplet structure. In this paper we pursue
such a direction, and in particular propose $B$ decays as a
possible avenue to uncover a host of such states. The $B$ meson
acts as a ``filter" for background suppression, which should be
compared with direct search of broad resonances in charm
fragmentation.

\section{\boldmath The Scalar $qq\bar q\bar q$ and $cq\bar q\bar q$ Multiplets}

Taking cue from the light hadron nonet, one considers the
{\boldmath $3^*$} under SU(3) formed by $qq$, i.e. $ud$, $us$ and
$ds$ where the flavors in the pair are distinct. Combining with
the {\boldmath $3$} of $\bar q\bar q$, one therefore gets an
{\boldmath $8$} plus a {\boldmath $1$}. The $\kappa$ state, if it
exists at all, is an isodoublet that can be written as $[ud\bar
d\bar s]^+$ and $[du\bar u\bar s]^0$, plus its antiparticle, where
we have suppressed all (anti-)symmetrizations of quark or
quark-antiquark pairings. The charged components for $a_0$ can be
written as $a_0^+ = [su\bar d\bar s]^+ = (a_0^-)^*$, with the
neutral component written as $a_0^0 = [s(n\bar n)_-\bar s]^0$ to
respect isospin ($(n\bar n)_\pm \equiv (u\bar u\pm d\bar
d)/\sqrt{2}$). We write $f_0 = [s(n\bar n)_+\bar s]^0$ which is
orthogonal to $a_0^0$ and is an isosinglet. This leaves $\sigma =
[ud\bar d\bar u]^0$ which is also a singlet. Although $f_0$ and
$\sigma$ (and for that matter, $a_0$) could mix, there is no
evidence for $f_0$ and $\sigma$ to deviate from the flavor content
we assign. The mass pattern supports isospin multiplets split by
their $s$ or $\bar s$ content.

\begin{figure}[t!]
%\smallskip\smallskip\smallskip
%\vskip-0.5cm
\centerline{ \hskip0.2cm
        {\epsfxsize2.7 in \epsffile{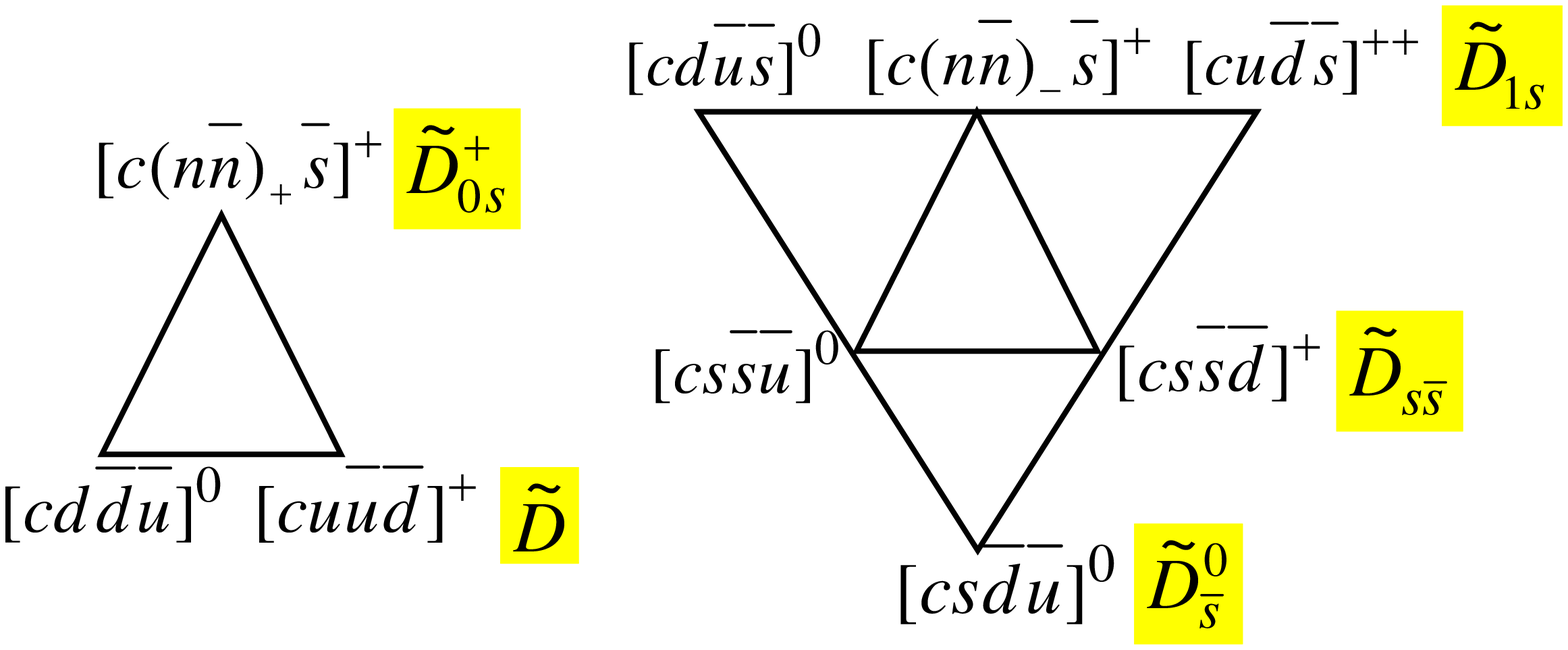}}
}
%\smallskip\smallskip\smallskip\smallskip
\vskip-2.5cm
\caption{ %
The {\boldmath $3^*$} and {\boldmath $6$} multiplets of $cq\bar
q\bar q$ mesons, where $cq = cu,\ cd,\ cs$, and $\bar q\bar q =
\bar d\bar s,\ \bar u\bar s,\ \bar u\bar d$.
} %
\label{fig:cqqq}
\end{figure}

Extending to the {\boldmath $3$} constructed from $cq$, one finds
the $cq\bar q\bar q$ 4-quark mesons form a {\boldmath $3^*$} and a
{\boldmath $6$}. Unlike Suzuki and Tuan \cite{Suzuki}, we think it
is the isospin multiplets that matter, with splittings between
multiplets determined more by $s$ or $\bar s$ content. Although
{\boldmath $3^*$}-{\boldmath $6$} splitting cannot be ruled out,
this is the assumption we shall follow. In the notation of Lipkin,
Suzuki and Tuan, the singlet of {\boldmath $3^*$} is denoted as
$\widetilde F_X^+$, while the triplet and singlet of {\boldmath
$6$} are denoted $\widetilde F_I$ and $\widetilde F_S^0$; we shall
denote these as $\widetilde D_{0s}^+ = [c(n\bar n)_+\bar s]^+$,
$\widetilde D_{1s} = ([cd\bar u\bar s]^0,\ [c(n\bar n)_-\bar
s]^+,\ [cu\bar d\bar s]^{++})$ and $\widetilde D_{\bar s}^0 =
[cs\bar u\bar d]^0$. Note that we have retained the tilde notation
of Lipkin for 4-quark states, but dropped $F$ since it is no
longer in use. Instead, we have extended the convention for
$D_s^+$ mesons, e.g. the $\widetilde D_{\bar s}$ state has $c$
{\it and} $s$ flavor, rather than $\bar s$! Following this
convention, and our assumption that the mass eigenstates are
determined by its (absence of) $s/\bar s$, $s\bar s$ content, we
denote the two doublets as $\widetilde D = ([cd\bar d\bar u]^0,\
[cu\bar u\bar d]^+)$ and $\widetilde D_{s\bar s} = ([cs\bar s\bar
u]^0,\ [cs\bar s\bar d]^+)$, where we assume ideal mixing.
These states, put in the form of {\boldmath $3^*$} and {\boldmath
$6$}, are illustrated in Fig. \ref{fig:cqqq}.

The $\widetilde D_{0s}^+$ state is an isosinglet, and can be
identified with the narrow $D_s(2320)$ state found by BaBar. We
shall argue that all other states decay hadronically. But we need
to first develop some picture on mass splittings. Besides the
$\widetilde D_{0s}^+$ state and the aforementioned $\widetilde
D_{\bar s}^0$ having $cs$ rather than $c\bar s$ flavor, another
notable state is the exotic doubly charged scalar $\widetilde
D_{1s}^{++}$.

\section{\boldmath Masses, Decay Modes and Widths}

This section should be viewed as providing only guesstimates.

As stressed in previous section, we treat isomultiplet masses
which are split by its $s$ or $\bar s$ quark content, rather than
work on SU(3) multiplet masses.
We shall ignore isospin splittings within an isomultiplet. We
therefore expect the mass ordering of $m_{\widetilde D} <
m_{\widetilde D_{Is}} \simeq m_{\widetilde D_{\bar s}} <
m_{\widetilde D_{s\bar s}}$, where $m_{\widetilde
D_{1s}}$-$m_{\widetilde D_{0s}}$ is susceptible to a possible
{\boldmath $6$}-{\boldmath $3^*$} splitting, which we ignore.

We illustrate with a picture of ``Naive Constituent Diquark Model"
for 4-quark mesons, in analogy to the usual Naive Quark Model of
normal mesons, just for counting purposes. Let us take
\begin{eqnarray}
m_{nn} &=& m_0, \ \ m_{sn} = m_0 + \Delta m_0,
\nonumber\\
m_{cn} &=& m, \ \ m_{cs} = m + \Delta m,
\label{eq:mqq}
\end{eqnarray}
where $n = u,\ d$, and we allow for $\delta m = \Delta m - \Delta
m_0$ to be nonzero, since the $cs$ and $ns$ pairing may have
different (QCD) dynamics. We then find that

\begin{eqnarray}
m_{\widetilde D} &\sim& m + m_0
 \sim \mbox{\rm 2100--2200 MeV},
\nonumber\\
m_{\widetilde D_{Is}} &\sim& m + m_0 + \Delta m_0
 \sim \mbox{\rm 2320 MeV},
\label{eq:mcqqq}\\
m_{\widetilde D_{\bar s}} &\sim& m + m_0 + \Delta m
 \sim \mbox{\rm 2320 MeV} + \delta m,
\nonumber\\
m_{\widetilde D_{s\bar s}} &\sim& m + m_0 + \Delta m_0 + \Delta m
 \sim \mbox{\rm 2500 MeV} + \delta m,
\nonumber
\end{eqnarray}
which realizes the above ordering, and allows $\widetilde D_{Is}$
to be heavier or lighter than $\widetilde D_{\bar s}$. The
assignment of $m_{\widetilde D_{Is}} \sim 2320$~MeV is clear. Let
us explain the other numbers.

Applying Eq. (\ref{eq:mqq}) to the $qq\bar q\bar q$ scalar nonet,
we find
$m_\sigma \sim 2m_0$,
$m_\kappa \sim 2m_0 + \Delta m_0$,
$m_{a_0,f_0} \sim 2m_0 + 2\Delta m_0$,
where the degeneracy of $a_0$ and $f_0$ is realized. The splitting
$m_\kappa - m_\sigma \sim 200$ MeV vs. $m_{a_0/f_0} - m_\kappa
\sim 180$ MeV is reasonable. We note, however, that the 4-quark
$\sigma$ state could be affected by chiral symmetry breaking which
leads to the pion being a pseudo-Goldstone boson. Mindful of this,
we take $m_0 \sim 300$~MeV, $\Delta m_0 \sim 180$~MeV. With these
numbers as input we arrive at the numerical suggestions in
Eq.~(\ref{eq:mcqqq}). Note that this numerology implies $m =
m_{cn} \sim 1800$~MeV, which is higher than the naive $m_c$
constituent mass. This is in contrast with $m_{nn} \sim m_n$ and
$m_{sn} \sim m_s$, and is one of the reasons behind our treatment
of $\delta m = \Delta m - \Delta m_0 = (m_{cs} - m_{cn}) - (m_{sn}
- m_{nn})$ as potentially nonzero. From this trend, we speculate
that $\delta m > 0$, hence $m_{\widetilde D_{\bar s}} \gtrsim
m_{\widetilde D_{Is}}$, and $m_{\widetilde D_{s\bar s}}$ is
probably above 2500~MeV.

With these masses, we now discuss decay modes.

By taking $m_{\widetilde D_{Is}} \sim 2320$~MeV, we identify the
isosinglet $\widetilde D_{0s}$ with the narrow BaBar state. Since
this is below $DK$ threshold, only the $D_s^+\pi^0$ isospin
violating decay is allowed, which is consistent with what is
observed. Electromagnetic decays are often on equal footing with
isospin violating decays, but BaBar seems to find these to be
subdominant. It is of interest to understand this.

The isospin-violating strong decay $\tilde D_{0s}^+\to D_s^+\pi^0$
can proceed via $\tilde D_{0s}^+\to D_s^+\eta(\eta')$ followed by
$\eta(\eta')-\pi^0$ mixing \cite{Cho}. We consider the flavor
mixing of $\eta_n$ and $\eta_s$ defined by
$\eta_n={1\over\sqrt{2}}(u\bar u+d\bar d)$ and $\eta_s=s\bar s$.
%, in analog to the wave functions
%of $\omega$ and $\phi$ in ideal mixing,
The wave functions of the $\eta$ and $\eta'$ are then given by
 \be
 \eta &=& \eta_n \cos\phi -\eta_s\sin\phi , \non \\
 \eta' &=& \eta_n\sin\phi +\eta_s\cos\phi.
 \en
A phenomenological analysis of many different experimental
processes indicates $\phi=39.3^\circ$ \cite{Kroll}.

The isospin violating term induced by the $u$ and $d$ quark mass
difference is
 \be
 {\cal L}={1\over 2}(m_d-m_u)(\bar uu-\bar dd).
 \en
The strong coupling for $\tilde D_{0s}^+\to D_s^+\pi^0$ then reads
 \be
 g_{\tilde D_{0s}^+D_s^+\pi^0}&=&\left( {m_d-m_u\over
 m_\eta^2-m_\pi^2}\sin^2\phi+{m_d-m_u\over
 m_{\eta'}^2-m_\pi^2}\cos^2\phi\right) \non \\
 &\times& {1\over 2}g_{\tilde
 D_{0s}D_s^+\eta_n} \la\pi^0|\bar uu-\bar
 dd|\eta_n\ra,
 \en
where $\la\pi^0|\cdots|\eta_n\ra$ can be evaluated
as~\cite{Ivanov}
 \be
 \la\pi^0|\bar uu-\bar dd|\eta_n\ra=\la\pi^0|\bar uu+\bar
 dd|\pi^0\ra=2v,
 \en
with $v=-2\la \bar qq\ra/f_\pi^2=m_{\pi}^2/( m_u+m_d)$. We shall
make two estimates of the strong coupling $g_{\tilde
D_{0s}D_s^+\eta_n}$. First, it can be extracted from the measured
width of $\kappa$ whose coupling is $g_{\kappa
K\pi}=\sqrt{3/2}\,g_0$ in terms of $g_{\tilde
D_{0s}D_s^+\eta_n}=g_0$ based on SU(4) symmetry. Using
$\Gamma_\kappa\approx 400$ MeV , we obtain $g_{\tilde
D_{0s}D_s^+\eta_n}\approx 4.4$ GeV and hence
 \be
 \Gamma(\tilde D_{0s}^+\to D_s^+\pi^0)\approx  11\,{\rm keV}.
 \en
Second, in the 2-quark model for $D_0^*$, the $P$-wave scalar, the
$D^*_0 D\pi$ coupling has been estimated in the framework of QCD
sum rules by two different methods, giving $g_{D^*_0 D\pi}=(6.3\pm
1.2)$ GeV and $(11.5\pm 4.0)$~GeV~\cite{Colangelo}. Since
$D^*_0\to D\pi$ is OZI suppressed relative to $\tilde D_{0s}\to
D_s^+\eta_n$, it is expected that $g_{\tilde
D_{0s}D_s^+\eta_n}>g_{D^*_0 D\pi}$. Taking $g_{\tilde
D_{0s}D_s^+\eta_n}=15$ GeV as a representative value, we are led
to $\Gamma(\tilde D_{0s}^+\to D_s^+\pi^0)= 130\,{\rm keV}$. In any
rate, it is fair to conclude that the width of $\tilde D_{0s}^+$
is smaller than 1~MeV and lies in between 10 and 100~keV.

For the radiative decay $\tilde D_{0s}^+\to D_s^+\gamma\gamma$,
three different contributions have been considered by
\cite{Suzuki}: (i) two-photon transition between four-quark
states, (ii) two-photon emission by pair annihilation, and (iii)
two photon emission from $\bar s$ quark. A naive estimate of
two-photon transition between $n\bar n$ and $n\bar n$ states are
\cite{Suzuki}
 \be
 \Gamma(\tilde D_{0s}^+\to D_s^+\gamma\gamma)\simeq {1\over
 12\pi}{\alpha^2\Delta ^5\over M^4},
 \en
where $\Delta =m_{\tilde D_{0s}}-m_{D_s}$ and $M$ is the
constituent quark mass of $u$ and $d$. Numerically, $\Gamma(\tilde
D_{0s}^+\to D_s^+\gamma\gamma) \simeq 0.48$ keV. For two-photon
emission by pair annihilation, it can proceed via $\tilde
D_{0s}^+\to D_s^+\eta$ followed by $\eta\to\gamma\gamma$. Current
algebra leads to~\cite{Suzuki}
 \be
 \Gamma(\tilde D_{0s}^+\to D_s^+\gamma\gamma)\approx {1\over
 192\pi^3}\left( {\alpha\over 8\pi f_\pi}\right)^2{\Delta ^9\over
 f_\pi^2 m_\eta^4},
 \en
giving $\sim 4\times 10^{-5}$ keV, which is further suppressed. We
conclude that $\tilde D_{0s}^+$ decay being dominated by the
isospin-violating strong decay is reasonable.

We have assumed that $m_{\widetilde D_{1s}} \simeq m_{\widetilde
D_{0s}}$ by resorting to quark content and some vague isospin
arguments. If this holds, since isospin splittings are rarely more
than 10~MeV, one is still below the $m_D+m_K$ threshold, but the
$\widetilde D_{1s} \to D_s^+\pi$ decay is now an allowed strong
decay, with a typical width of 100 MeV or more. Note that this
contains three modes: $\widetilde D_{1s}^{++} \to D_s^+\pi^+$,
$\widetilde D_{1s}^+ \to D_s^+\pi^0$, and $\widetilde D_{1s}^0 \to
D_s^+\pi^-$. The doubly charged scalar resonance would be
astounding.
We caution that there may still be some {\boldmath
$3^*$}-{\boldmath $6$} splitting, which would likely push the
{\boldmath $6$} higher than the {\boldmath $3^*$}~\cite{Suzuki}.
This could open up the $DK$ channels: $\widetilde D_{1s}^{++} \to
D^+K^+$, $\widetilde D_{1s}^+ \to D^+K^0$ and $D^0K^+$, and
$\widetilde D_{1s}^0 \to D^0K^0$. The doubly charged $D^+K^+$
scalar resonance would again be the most astounding.

The $\widetilde D$ doublet $\sim$ 2100--2200 MeV in mass would
decay into $D\pi$, just like a normal $D^{*}$ meson, but with flat
angular distributions, and a few 100 MeV in width. They should in
principle be distinct from the observed $D_1(2420)$ and
$D_2(2460)$ states since they are lower in mass. Since the $D\pi$
mass spectrum has been studied for some time, it may be difficult
to identify such scalar resonances, but efforts should be renewed.

The $\widetilde D_{s\bar s}$ is a second $D$-like doublet with
$s\bar s$ content, which pushes its mass above 2500 MeV, as we
have argued in Eq. (\ref{eq:mcqqq}). Thus, it is above $D_s^+\bar
K$ threshold of $\sim$ 2460 MeV. Besides the $\widetilde D_{s\bar
s} \to D_s^+\bar K$ channel, the $D\eta$ (and perhaps $D\pi$)
channel should be subdominant, with a slightly lower threshold,
and could provide a useful crosscheck.

We finally reach the $\widetilde D_{\bar s}$ exotic singlet with
both $c$ and $s$ flavor. We first offer another argument that
$\delta m > 0$ in Eq. (\ref{eq:mcqqq}). Suppose the opposite is
true. Since $\widetilde D_{0s}$ is below $DK$ threshold, so would
$\widetilde D_{\bar s}$ be below $D\bar K$ threshold. Inspection
of the $\widetilde D_{\bar s}$ quark flavor composition, it
contains all four distinct flavors, hence it can then undergo only
{\it weak} decay. Indeed, this was one of the original excitements
of Lipkin~\cite{Lipkin}, and of Suzuki and Tuan~\cite{Suzuki}. But
given that 20--25 years have elapsed, we find it highly unlikely
that we have missed a semi-stable, weakly decaying neutral
$D$-like meson at $\sim 2300$~MeV. Thus, we argue that
$m_{\widetilde D_{\bar s}}
> m_D + m_{\bar K}$, and one should search via
$\widetilde D_{\bar s} \to D^0\bar K^0$ or $D^+K^-$ with a strong
width of 100 MeV or more. Note the unusual pairing of $D$ and
$\bar K$ mesons.

The upshot of our discussion is that, BaBar found the single
narrow state which derives from its low mass, allowing it to decay
only via a suppressed isospin violating channel. All other $cq\bar
q\bar q$ scalar states undergo strong decay hence would be
considerably broader.

\section{\boldmath $B$ Decays as $\widetilde D_X$ Spectroscope}

All states except the $D_s(2320)$ already seen by BaBar would
likely be broad states. But to convince oneself of the 4-quark
nature, it would be necessary to uncover a major portion of the
full spectroscopy. The strongly decaying nature makes further
direct search in charm fragmentation not so optimistic. Instead,
we propose $B$ decays as a ``spectroscope" through which one may
search for such states with lower background.

\begin{figure}[b!]
%\smallskip\smallskip\smallskip
\vskip-0.3cm
\centerline{
\hskip3.8cm
        {\epsfxsize2.85 in \epsffile{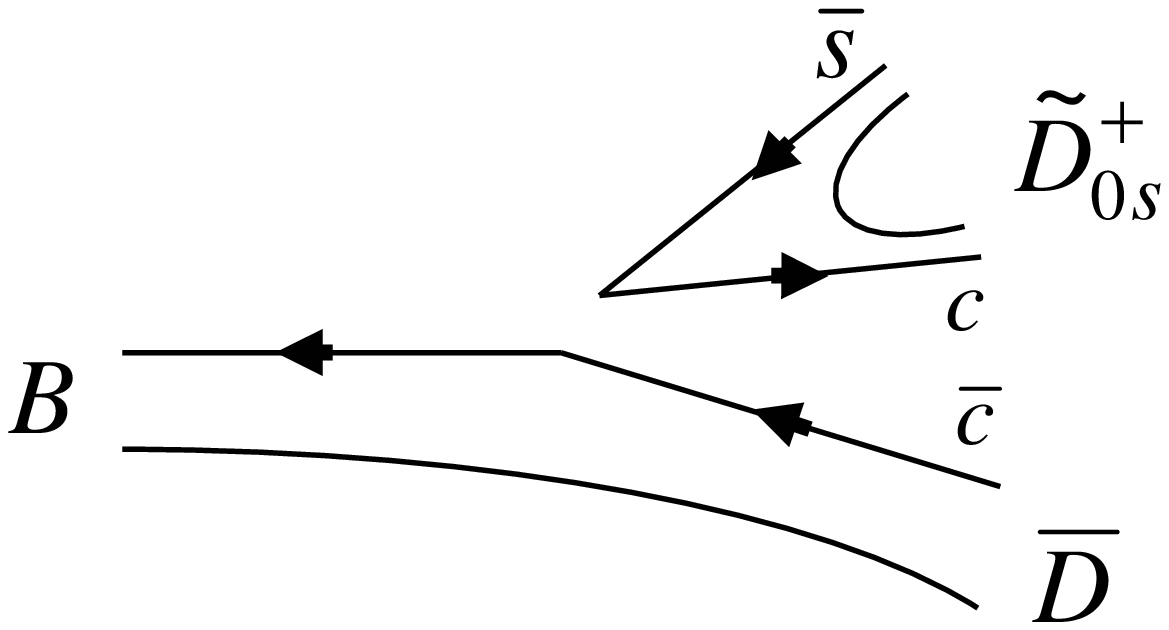}}
\hskip-3.3cm
        {\epsfxsize2.85 in \epsffile{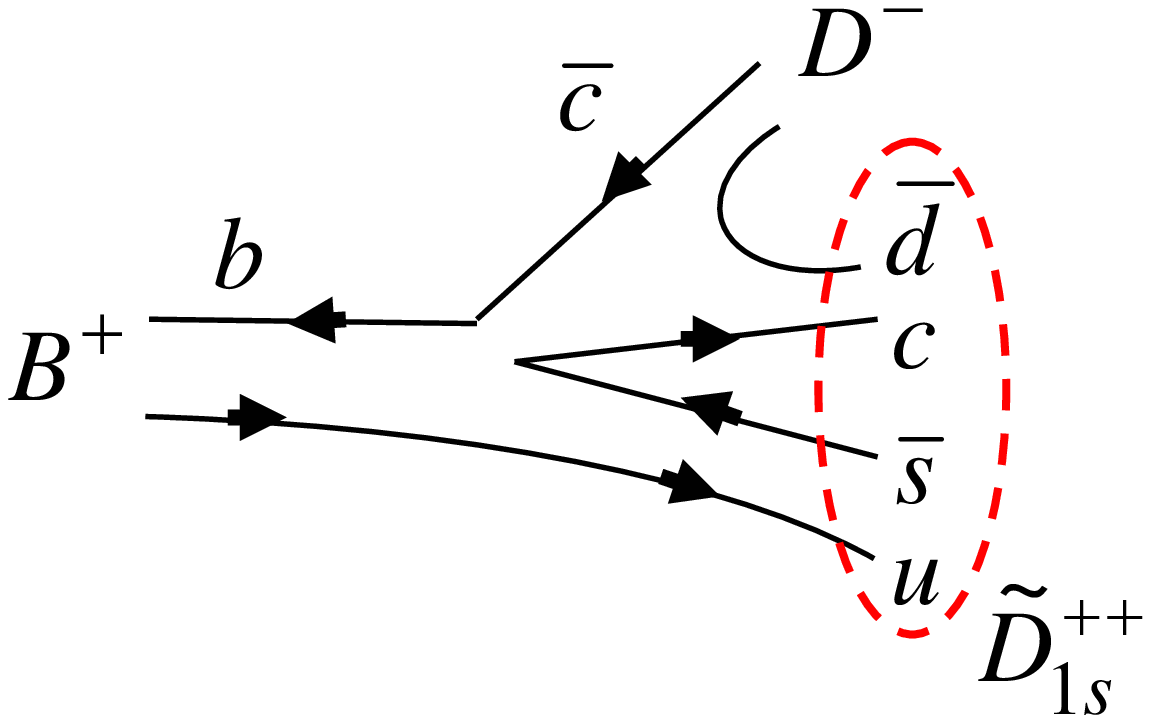}}
} \vskip-2.9cm \centerline{ \hskip3.8cm
        {\epsfxsize2.85 in \epsffile{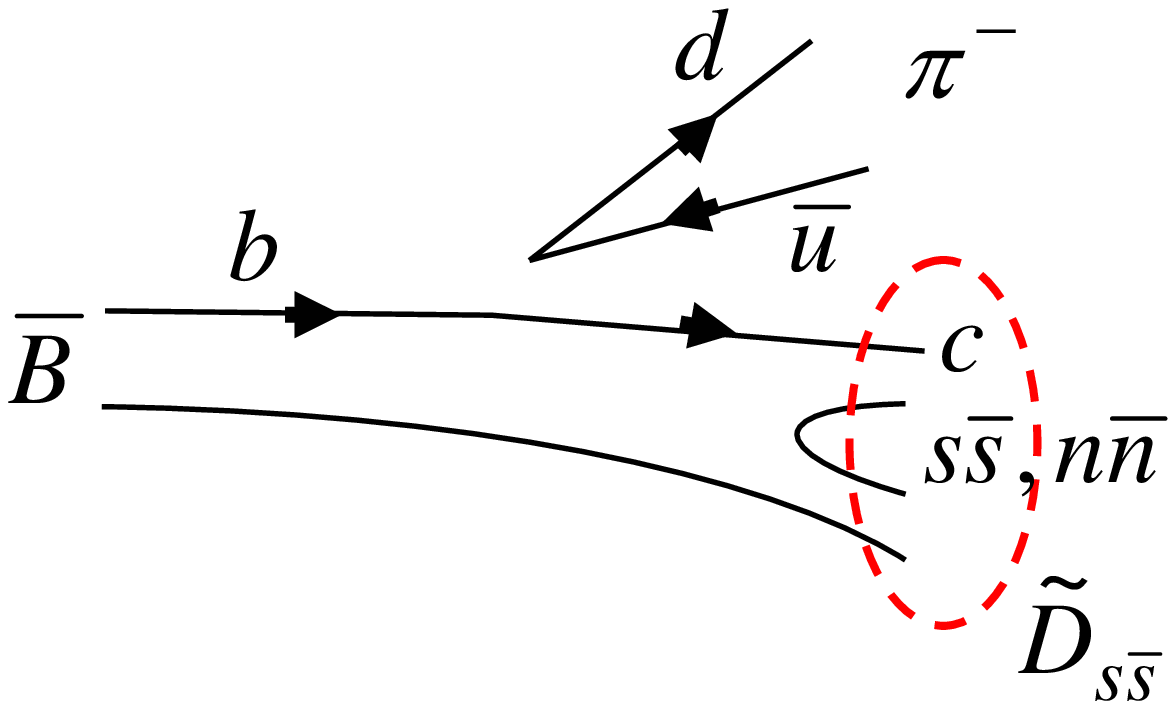}}
\hskip-3.3cm
        {\epsfxsize2.85 in \epsffile{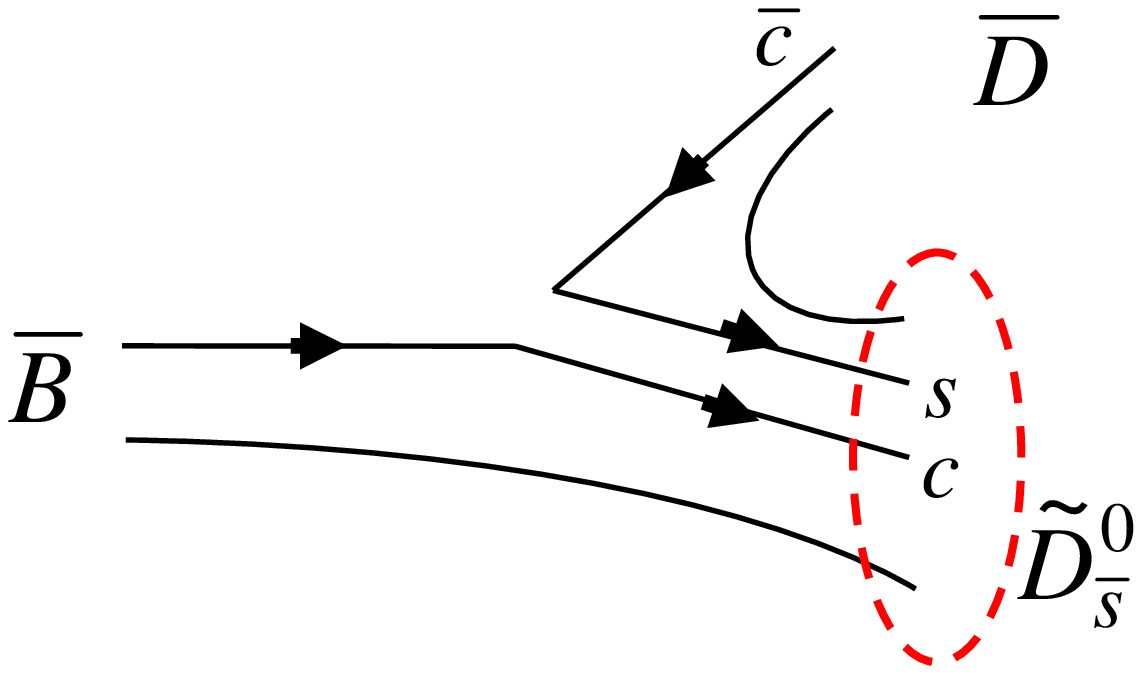}}
}
%\smallskip\smallskip\smallskip\smallskip
\vskip-3cm
\caption{ %
Diagrams for %
(a) $B\to \bar D\widetilde D_{0s}^+$, %
(b) $B^{+}\to D^-\widetilde D_{1s}^{++}$ ($B\to \bar D\widetilde D_{Is}$), %
(c) $\bar B\to \pi^-\widetilde D_{s\bar s}$, $\pi^-\widetilde D$, %
(d) $\bar B\to \bar D\widetilde D_{\bar s}^0$.
} %
\label{fig:btocqqq}
\end{figure}

We give 5 types of production processes, showing the richness.

\begin{description}
\item[{\boldmath $B\to \bar D^{(*)}\widetilde D_{0s}^+$}]:

The standard $\bar b\to c\bar c\bar s$ decay leads to $B\to \bar
DD_s^+$, which is factorization dominant. As illustrated in
Fig.~\ref{fig:btocqqq}(a), the $J^+$ vector current can produce a
scalar meson, which is not suppressed because of $c$ and light
quark mass imbalance. However, under factorization, it isprecisely
only the singly charged isosinglet state, $\widetilde D_{0s}^+$,
that can be produced.
Applying the equation of motion which leads
to
 \be
  m_{a_0}^2f_{a_0^-} &=& i(m_d-m_u)\la a_0^-|\bar du|0\ra, \non \\
 m_{\widetilde D_{0s}}^2f_{\widetilde D_{0s}} &=&
  i(m_c-m_s)\la \widetilde D_{0s}|\bar cs|0\ra,
 \en
and assuming $\la \widetilde D_{0s}|\bar cs|0\ra\approx \la
a_0^-|\bar du|0\ra$, it follows that $f_{\widetilde D_{0s}}\approx
67\pm 13$ MeV for $m(\widetilde D_{0s})\approx 2.32$ GeV and
$f_{a_0^\pm}=1.1\pm 0.2$ MeV obtained from finite-energy sum
rules~\cite{Maltman}. For $f_{D_s}\sim 230$ MeV, this means that
the production rate of $B\to \bar D \widetilde D_{0s}^+$ is
smaller than that of $B\to \bar D D_{s}^+$ by one order of
magnitude.  It is important to confirm the $D_s^+(2320)$ state
recoiling against a $\bar D$ meson in $B$ decay.
\item[{\boldmath $B^+\to D^{(*)-}\widetilde D_{1s}^{++}$
 ($B\to \bar D^{(*)}\widetilde D_{Is}$)}]:

To overcome the limitation of Fig.~\ref{fig:btocqqq}(a) to singly
charged isosinglet state under factorization, we construct a
process whereby the exotic doubly charged $\widetilde D_{1s}^{++}$
state can be produced. This is illustrated in
Fig.~\ref{fig:btocqqq}(b), where the popping of a $d\bar d$ pair
between $c\bar c$ allows $B^+$ to decay to a $D^{(*)-}$ plus the
$\widetilde D_{1s}^{++}$. The latter can be searched for in $
D_s^+\pi^+$, or possibly $D^+K^+$ channels. It would be
astonishing if a doubly charged resonance is found. Note that
Fig.~\ref{fig:btocqqq}(b) provides a generic mechanism for $B\to
\bar D\widetilde D_{s}$, i.e. all four $\widetilde D_{s}$ states
can be produced this way, with appropriate pairing with a $\bar D$
meson. For example, the neutral state can be produced via $B^0\to
\bar D^0\widetilde D_{1s}^0$, which can be searched for via
$\widetilde D_{1s}^0\to D_s^+\pi^-$, or perhaps $D^0K^0$ if
allowed.
\item[{\boldmath $\bar B\to \pi^-\widetilde D_{s\bar s}$,
$\pi^-\widetilde D$}]:

We illustrate in Fig.~\ref{fig:btocqqq}(c) a second type of
factorized production process, via $b\to c\bar ud$. A $\pi^-$ is
emitted in standard way, but the $b\to c$ current excites the
effective popping of an $s\bar s$ pair in the recoil system, and a
$\widetilde D_{s\bar s}$ state can be formed, which decays to
$D_s^+\bar K$ or perhaps $D\eta$. Likewise, but perhaps more
difficult to disentangle experimentally, the $\widetilde D$ state
can be produced, which would decay to $D\pi$.
\item[{\boldmath $\bar B\to \bar D^{(*)}\widetilde D_{\bar s}$}]:

To see how the exotic $\widetilde D_{\bar s}$ state can be
produced, we reverse all quark directions in
Fig.~\ref{fig:btocqqq}(a) and reshuffle the $s$ quark to pair with
the $c$ quark from $b\to c$ transition, as illustrated in
Fig.~\ref{fig:btocqqq}(d). It is not clear whether Nature holds
sufficient dynamics for this, but it is best to resort to data.
Note that the $\widetilde D_{\bar s}$ leads to ``wrong" pairings
of $D^0\bar K^0$ and $D^+K^-$ compared to usual $B$ decay.
\item[{\boldmath $\bar B^0\to \bar K\widetilde D_{s}$,
$K^0\widetilde D_{\bar s}^0$}]:

Finally, we turn to nonfactorized processes that may also produce
$cq\bar q\bar q$ states. The process $\bar B^0 \to K^- D_s^+$ has
been observed by Belle \cite{BelleKDs} and BaBar \cite{BaBarKDs}
at a rate that cannot come from factorized amplitudes, but arise
from either final state rescattering, or from annihilation
diagrams. Either way, annihilation, exchange of constituents, and
$q\bar q$ popping can provide further avenues for 4-quark state
search. For example, based on $s\bar uc\bar s$ configuration
already present in $\bar B^0\to K^-D_s^+$ final state, popping of
$d\bar d$ or $u\bar u$ pairs can lead to $\bar B^0\to
K^-\widetilde D_{Is}^+$, $\bar K^0 \widetilde D_{1s}^0$, as well
as the $K^0 \widetilde D_{\bar s}^0$ final states.
Thus, one not only should see the $D_s(2320) \to D_s^+\pi^0$ state
recoiling against a single $K^{(*)-}$ (with overlap of the narrow
and the broad state), but perhaps also search for $\bar B^0\to
K_S[D_s^+\pi^-]$ ($K_S[D^0K^0]$) and $\bar B^0\to K_S[D^0\bar
K^0]$, $K_S[D^+\bar K^-]$. Besides the distinct $D^+\bar K^-$
channel, for $D^0K_S$ one in principle can also use mass
separation.
\end{description}

Our discussion has only been illustrative rather than exhaustive.
The main point is that $B$ decays may act as a ``filter" through
$B$ reconstruction. In comparison, although the $D_s(2320)$ state
was discovered in charm fragmentation of $e^+e^-\to c\bar c$, to
find broad resonances through such processes would be much more
difficult.

\section{\boldmath Exotic Charmonia and Slow $J/\psi$ from $B$ Decay?}

It is natural to extend from $cq\bar q\bar q$ to $cq\bar q\bar c$
scalar states, which form an {\boldmath $8 \oplus 1$}. Continuing
our main assumption, the octet consists of an isotriplet and
isosinglet composed of $[nc\bar c\bar n]$, plus two isodoublets
$[nc\bar c\bar s]$ and its charge conjugate. The heavy singlet has
$[sc\bar c\bar s]^0$ structure. If we extrapolate from the naive
model of Eq. (\ref{eq:mqq}), we find
\begin{eqnarray}
m_{nc\bar c\bar n} &\sim& \mbox{\rm 3600 MeV} \lesssim m_{D\bar D},
\nonumber \\
m_{nc\bar c\bar s} &\sim& \mbox{\rm 3800 MeV} \lesssim m_{D_s\bar
D},
\nonumber \\
m_{sc\bar c\bar s} &\sim& \mbox{\rm 4000 MeV} \lesssim m_{D_s\bar
D_s},
\label{eq:mcqqc}
\end{eqnarray}
although whether $[sc\bar c\bar s]^0$ is still below $D_s\bar D_s$
threshold is more dubious.

From Eq. (\ref{eq:mcqqc}) one can infer the dominant decay modes.
If the isosinglet $[nc\bar c\bar n]_0^0$ is below $D\bar D$, the
dominant decay would be $[nc\bar c\bar n]_0^0\to \eta_c\eta$,
since it is clearly above the $\eta_c\eta$ threshold. For the
isotriplet, $[nc\bar c\bar n]_1\to \eta_c\pi$ should be the
leading decay, but the $[nc\bar c\bar n]_1\to J/\psi \rho^*\to
J/\psi \pi\pi$ decay, with $\pi\pi$ peaking towards $m_\rho$,
could be substantial. If the isodoublet $[nc\bar c\bar s]$ is
below $D_s\bar D$ threshold, the leading decays may be $\eta_c K$,
or $J/\psi K\pi$ with $m_{K\pi}$ peaked towards $m_K^*$ (perhaps
less prominent than $\rho$ case because of smaller $K^*$ width).

We are intrigued to stress that there may be some bearing for this
already, in the observed ``lump" for slow momentum $J/\psi$s below
$p^*_{J/\psi} \lesssim 1$ GeV observed in inclusive $B$ decay.
Such effect has been observed by all three
experiments~\cite{slowJ}, and can only be of hadronic origin. The
scenario of $B\to K^{(*)}+[nc\bar c\bar n]_1 \to K+J/\psi
(\pi\pi)_{\rho^*}$ and $B\to (\pi,\ \rho)+[nc\bar c\bar s] \to
K+J/\psi (K\pi)_{K^*}$ provide interesting possibilities whereby
$J/\psi$ is forced slow by $m_{\pi\pi}$ ($m_{K\pi}$) peaking
towards $\rho$ ($K^*$) mass. But since in general the
$K^{(*)}\eta_c\eta$ or $(\pi,\ \rho)\eta_c K$ modes should be
dominant, one should search for these modes as crosschecks. Not
only one has interest to understand what is behind the slow
$J/\psi$ bump in $B$ decay, one may possibly uncover $J/\psi
\pi\pi$, $\eta_c\eta$, $\eta_c\pi$, or $J/\psi K\pi$, $\eta_c K$
4-quark resonances.

Finally, if the $[sc\bar c\bar s]^0$ configuration resonates, it
may lead to $B\to K\phi J/\psi$ (observed by CLEO~\cite{JphiK})
and $B\to K\eta^{(\prime)} \eta_c$, which also should be studied.

Incidentally, we note that in principle $c\bar c$ could annihilate
via a single virtual gluon, i.e. $c\bar c \to g^* \to q\bar q$. It
would be interesting to see if the $B\to KK_SK^-\pi^+$ decay,
which Belle claimed~\cite{Ketac2S} to uncoverthe $\eta_c(2S)$
state in $K_SK^-\pi^+$ decay, could contain some information on a
possible $[nc\bar c\bar n]$ state, as the $\eta_c(2S)$ mass found
at 3654 MeV seems on the high side (by $\sim 100$ MeV compared to
potential model expectations, but could be consistent with the
$[nc\bar c\bar n]$ states.

We should also stress that exotic $0^+$ charmonia could also be
searched for directly in charm fragmentation, e.g. in the recoil
system of $e^+e^-\to J/\psi +X$ at B Factory energies. It may in
fact shed light on the rather mysterious $e^+e^-\to cc\bar c\bar
c$ production observed by Belle.

\section{Discussion and Conclusion}

There is one potential problem for the 4-quark interpretation of
$D_s(2320)$ observed by BaBar: There is also hint~\cite{BaBarDs0}
for a 2460 MeV state that decays to $D_s^*\pi^0$, i.e. in the
$D_s^+\pi^0\gamma$ final state. The existence of this state was
not yet conclusive in the BaBar paper, but if it holds as another
resonance, it cannot be of $0^+$ quantum number. The naive guess
would be $1^+$. If so, one interpretation of the $D_s(2320)$ and
``$D_s(2460)$" pair would be the $j = 1/2$ $s$-spin doublet of
$0^+$ and $1^+$. The question then gets translated into why these
states seem considerably lower in mass than expected. For
instance, the paper by Di~Piero and Eichten \cite{QM} gives the
masses at 2490 MeV and 2600 MeV, respectively, and one usually
expects these states to be broad. If the $0^+$ state moves below
$DK$ threshold, one ends up with the isospin violating
$D_s^+\pi^0$ decay as dominant mode. But for the $1^+$ state,
further mystery is why its rate is not dominated by $D_s(2460) \to
D_s^+\pi\pi$, an allowed strong decay; $D_s(2460) \to D_s^*\pi^0$
decay still would violate isospin. The width observed by BaBar for
the 2460 MeV state is barely larger than resolution. We think it
is far fetched at present to consider $1^+$ states composed of
$cq\bar q\bar q$ (in principle one could also consider $cg\bar q$
``hybrids"), and again the relative narrowness would come into
question. We look forward to clear establishment of the
$D_s^{*+}\pi^0$ state at 2460~MeV to clarify the situation.

In conclusion,
starting from the observed narrow $D_s(2320)$ state by BaBar,
together with the possibility of an emerging light scalar nonet,
we have explored the possible interpretation via $cq\bar q\bar q$
4-quark states, which could naturally be below $DK$ threshold.
We give a naive picture of the states, their masses, and infer the
corresponding decay modes. Only the $D_s(2320)$ state, identified
as the $\widetilde D_{0s}^+$ isosinglet with $[c(n\bar n)_+\bar
s]^+$ composition, decays dominantly by isospin violating decay
and is narrow. The remaining triplet, two doublets, and a heavy
singlet all decay hadronically, hence broad. The decay signatures
are distinct. Particularly noteworthy are resonances in the doubly
charged $D_s^+\pi^+$ ($D^+K^+$), and wrong pairing $D^+K^-$
channels.
We propose a host of $B$ decays as possible ``spectroscopes" to
search for these exotic resonances, and give many explicit
channels for further study. The hadronic width of these particles
may hamper the analogous program for direct search in charm
fragmentation.
We extend the picture to include $qc\bar c\bar q$ states, which
may lead to exotic $0^+$ charmonia resonances that could be behind
the slow $J/\psi$ excess in $B$ decay, and may shed light on
$e^+e^- \to cc\bar c\bar c$ production.
It is clear that only by establishing a clear spectroscopy of at
least half the multiplets, can one start to put some faith in the
$cq\bar q\bar q$ (or $qc\bar c\bar q$) 4-quark meson scenario.

\vskip 0.2cm
\noindent  %{\bf Acknowledgement}.\ \
WSH wishes to thank M. Yamauchi for bringing the subject to his
attention, and A. Bondar, T. Browder, A.~Drutskoy, H.C. Huang,
J.~Mueller, B.~Yabsley for discussions. This work is supported in
part by grants from NSC 91-2112-M-001-038, 91-2112-M-002-027,
 the MOE CosPA Project,
and %the BCP Topical Program of
NCTS.

\vskip 0.3cm \noindent  {\bf Note Added}.

\vskip-0.055cm \noindent After this work was completed, we noticed
the appearance of the related works by
R.N. Cahn, J.D. Jackson, hep-ph/0305012, and
T. Barnes, F.E. Close, H.J.Lipkin, hep-ph/0305025.

\end{document}